\documentclass[aps,pre,twocolumn,superscriptaddress,showpacs]{revtex4-1}
\usepackage{graphicx}
\usepackage{dcolumn}
\usepackage{bm}

\usepackage{natbib}
\usepackage[
text={7in,9.5in},centering,
]{geometry}

\usepackage{epsfig}
\usepackage{amsmath}
\usepackage{amssymb}
\usepackage{textcomp}

\begin{document}


\title{
Structural stability  of interaction networks against negative external fields.
}


\author{S. Yoon}
\affiliation{Department of Physics $\&$ I3N, University of Aveiro, 3810-193 Aveiro, Portugal}
\author{A. V. Goltsev}
\affiliation{Department of Physics $\&$ I3N, University of Aveiro, 3810-193 Aveiro, Portugal}
\affiliation{A. F. Ioffe Physico-Technical Institute, 194021 St. Petersburg, Russia}
\author{J. F. F. Mendes}
\affiliation{Department of Physics $\&$ I3N, University of Aveiro, 3810-193 Aveiro, Portugal}




\begin{abstract}
We explore structural stability of weighted and unweighted networks of positively interacting agents against a negative external field.
We study how the agents support the activity of each other to confront the negative field, which suppresses the activity of agents and can lead to a collapse of the whole network. The competition  between the interactions and the field shape the structure of stable states of the system. In unweighted networks (uniform interactions)
the stable states have the structure of $k$-cores of the interaction network.
The interplay between the topology and the distribution of weights (heterogeneous interactions) impacts strongly the structural stability against a negative field, especially in the case of fat-tailed distributions of weights.
We show that apart critical slowing down there is also a critical change in the system structure that precedes the network collapse. The  change can serve as early warning of the critical transition. In order to characterize changes of network structure we develop a method  based on statistical analysis of the $k$-core organization and so-called `corona' clusters belonging to the $k$-cores.
\end{abstract}

\maketitle


\section{Introduction}
\label{introduction}

The impact of negative external factors such as catastrophic environmental
changes, anthropogenic  or media pressure
on technological, biological, and social complex networks can lead to collapse of the systems when interactions between subjects forming the systems cannot  resist anymore the factors \cite{barnosky2012approaching,rothman2017thresholds}.
In biology,  stability of ecological networks against negative external factors
is provided in a large extent by mutualistic interactions between species \cite{may1972will}.
Mutualism is a relationship between organisms of different species when each individual benefits from the activity of the other. Mutualistic networks, which represent mutualism,
are of particular interest in recent studies
of ecosystems \cite{thompson2006mutualistic}.
Another example of mutualistic system is a social group linked by common religious, cultural, or political interests. This group
can be destroyed by a negative influence of media while a strong mutualistic (attractive) interaction can provide stability of the system  and confronts
the negative external pressure. It is well recognized that the structure plays a very important role in the robustness of complex systems against errors \cite{albert2000error,albert2002statistical,dorogovtsev2002evolution,newman2003structure,dorogovtsev2008critical}, the stabilization of  ecosystems against habitat destruction, alien species introduction, climate change, or pollution \cite{thebault2010stability,dunne2002network,memmott2004tolerance,jordano2006ecological,lever2014sudden,rohr2014structural}, and  resilience of social networks \cite{garcia2013social,torok2017cascading}.
The big questions in complex systems science are what causes some systems to collapse, how to predict the approach to the tipping point, what is the role of network structure in stability of real complex systems \cite{scheffer2009early,scheffer2012anticipating}.

One important characteristics of network structure
is the network cohesiveness.
Seidman introduced a so-called `$k$-core' in order to characterize the cohesion in social networks \cite{seidman1983network}. The $k$-core is the largest subgraph whose all vertices have, at least, $k$ nearest neighbors. The $k$-core is obtained by a pruning process as follows. Remove vertices with degree less than $k$. If there are vertices which have degree less than $k$ as a consequence of the previous removal, these vertices are also pruned from the network until there is no more vertex to be removed. The final maximal subgraph with the sequential pruning process is the $k$-core.  Any complex network can be represented as a set of nested $k$-cores with the core index $k$ running from 2 to $k_h$, where the index $k_h$ characterizes the highest $k$-core. Note that $2$-core includes $3$-core as subgraph. In turn, $3$-core includes $4$-core, and so on.
%
%
%
The $k$-core organization of a complex network is determined by its topological structure \cite{dorogovtsev2006k,dorogovtsev2006kpercolation,goltsev2006k}.
The highest $k$-core is characterized by the maximum  core index $k_h$, which is topological invariant of the network.
Analysis of $k$-cores  was used to characterize the structure of various real complex networks \cite{dorogovtsev2006k,dorogovtsev2006kpercolation,dorogovtsev2008critical,Alvarez2008k-core,kitsak2010identification} including plant-pollinator mutualistic networks \cite{jordano2006ecological,Fang2012}, social networks \cite{garcia2013social}, biological networks \cite{hagmann2008mapping,silva2015amyloid}, and many other networks. Since the $k$-cores represent the most connected part of a network, one would expect that the $k$-core organization might play an important role in the structural stability of real complex network against damages and negative external factors.

The researches mentioned above considered interaction networks as unweighted networks.
However, many real complex systems are best described by weighted networks where weights represent, for example, strengths of interactions \cite{albert2002statistical,newman2003structure,Barrat04,bullmore2009complex,van2016food}.  Structural properties of weighted networks need a special consideration, which takes into account both topological organization of networks and weights distributions \cite{Barrat04,csermely2013structure}. Fat-tailed distributions of weights are of special interest because they were found in many real systems such as neuronal networks and ecosystems.  At the present time, the understanding of the impact of the network topology and weight distribution on dynamics and stability of complex networks is still elusive (see, for example, a recent discussion of structural stability of food webs \cite{van2016food} and mutualistic systems \cite{rohr2014structural}). New methods of structural analysis of weighted networks are also necessary.

In this paper, we explore the role of topology and the heterogeneity of interactions in the structural stability of networks of positively interacting agents subjected to a negative external field, which suppresses the  activity of the agents.
We study how positively interacting agents support each other to confront the negative field and the role of the $k$-core organization in the structural stability of the interacting system. In our approach we understand the structural stability as the existence of a giant connected component of the network of active agents stable against perturbation.
In the case of unweighted  (uniform interactions) networks,
we demonstrate that the tipping point of network collapse caused by a strong negative field is determined by the highest $k$-core.
In weighted networks (heterogeneous interactions), we study the interplay between the topology and the distribution of weights in the structural stability against a negative field.
We also develop a new method of structural analysis based on statistical analysis of so-called `corona' clusters belonging to $k$-cores. This method allows us to reveal structural changes in the $k$-core organization when increasing the negative external field and allow to predict collapse of weighted and unweighted networks.
Structural stability of some real networks against negative external fields is also discussed.


\section{Model}
\label{model}
Let us consider a system of $N$ interacting agents. Every agent $i$, $i=1,2,\dots, N$, can be either in active or inactive state. If agent $i$ is active then the parameter $x_i$ is 1, otherwise $x_i=0$. We characterize the energy $E$ of the system as follows,
\begin{equation}
\label{eq:1}
  E  = - \frac{1}{2} \sum_{ij} w_{ij} A_{ij} x_{i} x_{j} - \sum_{i=1}^{N} U_{i} x_{i}.
\end{equation}
Here, the structure of the interaction network
is determined by the adjacency matrix  $A_{ij}$ with entries $A_{ij}=1$ if agent $i$ acts on
agent $j$, and $A_{ij}=0$, otherwise. Moreover, $A_{ii}=0$ for all $i=1,2,\dots, N$. The edge weight  $w_{ij}$ determines the strength of action of agent $i$ on agent $j$. In general case, the matrices $A_{ij}$ and $w_{ij}$ can be asymmetric, i.e., directed or bipartite. The parameter $U_i$ is an external field acting on agent $i$. In the framework of the model, positive interactions $w_{ij} >0$ stabilize the system of interacting agents. Positive interactions can represent mutualistic interactions between agents when agent $j$ benefits from the presence of agent $i$.
Negative weights $w_{ij} < 0$ represent antagonistic interactions between agents.  Negative fields $U_i <0$ represent negative external factors that  deactivate agents, while positive fields $U_i > 0$ attract agents $i$ into the system. In the framework of the model, we understand the structural stability of the system as the existence of the giant connected component in the ground state.
This condition assumes that there is a finite fraction of interconnected active agents in the state and stability against weak fluctuations in the number of active agents.

\section{Structural stability of unweighted  networks}
\label{unweighted nets}
Let us study structural stability of an unweighted undirected network of positively interacting agents, i.e., $A_{ij}=A_{ji}$ and  $w_{ij}=w >0$.
The external field is uniform and negative,  $U_i = U <0$. We aim to show that the negative external field shapes the structure of the ground state of the system Eq. (\ref{eq:1}). The ground state is the $k$-core, if it exists, formed by active agents in the considered network. In this state the core index $k$ is
\begin{equation}
 k=\Bigl[ \frac{|U|}{w} \Bigr] +1.
\label{eq:2}
\end{equation}
Here, $[x]$ denotes the integer part of a real number $x$. It is convenient to use the following representation,
\begin{equation}
U/w = - (k-1 +\delta),
\label{eq:3}
\end{equation}
where $\delta \in (0,1]$. Note that only integer numbers $k \geq 1$ correspond to  negative $U$. 
Choosing the interaction energy $w$ as the energy unit, we write the energy $E$ in Eq. (\ref{eq:1}) in a form,
\begin{equation}
\label{eq:4}
  E  = - \frac{1}{2} \sum_{ij} A_{ij} x_{i} x_{j} + \sum_{i=1}^{N} (k-1 +\delta) x_{i}.
\end{equation}
The energy of activation ($x_i =1$) or deactivation  ($x_i =0$) of agent $i$ is
\begin{equation}
\label{eq:5}
e(i)  = - \Bigl(\sum_{j} A_{ji}x_{j} - k +1 -\delta \Bigr)x_i.
\end{equation}
At first we consider the case $k=1$, i.e., $U=-\delta$. Agents, which have no interaction with other agents (isolated nodes), are inactive ($x_i=0$) in the ground state because if they are active then the total energy is increased by value $\delta$ per isolated agent.
For simplicity, throughout this paper, we assume that there are no isolated agents in the initial state.
If all interacting agents are active, then the energy per agent is
\begin{equation}
\label{eq:6}
  \frac{E(k=1)}{N}  = - \frac{L}{N} + \delta = - \frac{1}{2} \langle q \rangle  + \delta
\end{equation}
where $L=\frac{1}{2} \langle q \rangle N$ is the total number of edges (interactions) in the network
and  $\langle q \rangle = \sum_{q} q P(q)$ is the mean degree in the network with the degree distribution $P(q)$.
If $\delta < 1/2$, then in the ground state we have $x_i=1$ for all $i=1,2,\dots, N$.
The main contribution into Eq. (\ref{eq:6}) is given by a giant percolating component formed by the active agents. There are also disjoint finite clusters of interacting agents.
A pair of interacting agents is stable against negative external factors because it has a negative energy $-1+2 \delta < 0$ at $\delta <1/2$. If  $\delta $ increases above $1/2$, then  disjoint pairs and small finite clusters of interacting agents become inactive, but large clusters  may be still active. For example, a tree-like cluster of size $n$ has $n-1$ edges and a positive energy $-n+1+ n \delta$ if $ \delta > 1-1/n$. Therefore agents in this cluster are deactivated by the field to decrease the total energy.  However, agents belonging to a cluster of size $n>1/(1-\delta)$ are active. Note that the giant percolating component formed by interacting agents is stable against the negative field at any $\delta \in (0,1]$. When $\delta \rightarrow 1$, all agents in finite clusters  are deactivated.

\begin{figure}[!ht]
 \centering
 \includegraphics[width=80mm,angle=0.]{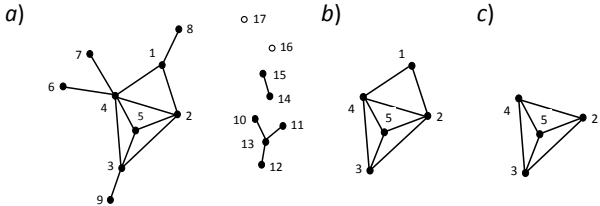}
 \caption{Example of an unweighted network of interacting agents in negative external field, Eq. (\ref{eq:3}). (a) In the case $k=1$ ($U=-\delta$, where $\delta < 1/2$), all agents forming finite clusters (nodes 10-13 and 14-15) and a `giant' connected component (nodes 1-9 ) are active, while isolated agents 16 and 17 are inactive (open circles). (b) At $k=2$ ($U=-1-\delta$), only agents forming $2$-core are active. (c) At $k=3$ ($U=-2-\delta$),  only agents forming $3$-core are active. }
 \label{figNET}
\end{figure}

Let us consider the case $k-1 < |U/w| <k$ at $k \geq 2$. In the initial state all agents are active, i.e., $x_i =1$ at all $i=1,2, \dots$. Therefore, according to Eq. (\ref{eq:5}), agent $i$ contributes an energy,
\begin{equation}
\label{eq:7}
e(i)  = - (q_i - k +1 -\delta),
\end{equation}
into the total energy $E$. The energy $e(i)$ is positive if
degree $q_i$, i.e., the number of agents with which agent $i$ interacts, is smaller than $k-1$.
Therefore, the total energy $E$ decreases if agent $i$ with degree $q_i \leq k-1$ becomes inactive. We put $x_i =0$ for this agent. Then, using Eq. (\ref{eq:5}), we recalculate the contributions of remaining agents and again remove all agents having less than $k$ remaining active partners. This pruning algorithm converges to a state with a minimum energy. This state is the $k$-core, if it exists, defined in Sec. \ref{introduction}.  The $k$-core state formed by interacting  agents is stable against both removal and addition of other agents by construction.

The energy of the $k$-core state is
\begin{equation}
\label{eq:8}
  \frac{E_k}{N}  = {-} \frac{L_k}{N} {+} (k{-}1 {+} \delta) M_k= - \frac{1}{2} \Bigl[\langle q \rangle_k  - 2(k-1+\delta)\Bigr]  M_k
\end{equation}
where $L_k$ and $\langle q \rangle_k$ are the number of edges and the mean degree in the $k$-core, respectively. $M_k$ is the fraction of nodes in the $k$-core, $L_k /N = \langle q \rangle_k M_k /2$. Is the $k$-core the ground or metastable state? In order to answer this question, we consider the state with  $x_i=0$ and the total energy $E=0$. This inactive state is stable against activation of a small fraction of randomly chosen agents.
If the total energy $E_k < 0$, then the $k$-core is the ground state and the inactive state $E=0$ is metastable. If $E_k > 0$, then the $k$-core is metastable and the inactive state $E=0$ is the ground state.

\begin{figure}[!ht]
 \centering
 \includegraphics[width=80mm,height=50mm,scale=0.7]{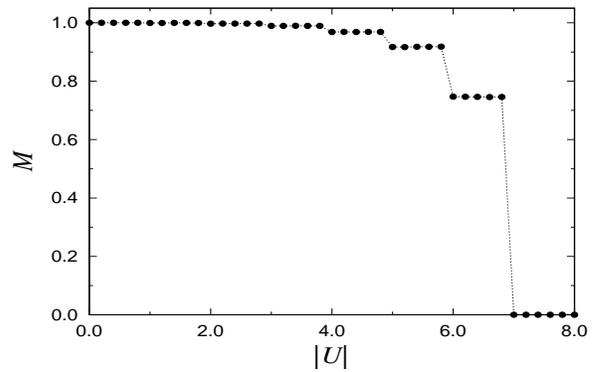}
 \caption{Fraction of active agents $M$  versus the field magnitude $|U|$ in the ground state of the ER network of size $N=10^4$ and the mean degree $\langle q \rangle =10$. Jumps occur at $|U|=2,3,\dots,7$. Results are averaged over 100 realizations.}
 \label{MvU}
\end{figure}


As an example we consider the model Eq. (\ref{eq:4}) on a classical random graph such as the Erd\H{o}s-R\'{e}nyi (ER) random network, which is a representative model of random uncorrelated networks with finite second moment of the degree distribution. Figure \ref{MvU} displays the dependence of the fraction $M$ of active agents,
\begin{equation}
\label{eq:9}
  M \equiv \frac{1}{N} \sum_i x_i,
\end{equation}
on $|U|$ in the ground state of the ER random network. When  increasing  $|U|$ the fraction $M$ of active agents  undergoes abrupt jumps corresponding to transition from $k$-core to $(k+1)$-core state. Above the tipping point  $|U_c| = 7$, which corresponds to the highest $k$-core with the core index $k=k_h$ ($k_h =7$ for ER network with $\langle q \rangle =10$), there is no active agent in the ground state and, therefore, $M=0$ at $|U| > |U_c|$.
Table \ref{table1} represents our numerical results at $\delta =0.001$.
 One can see that at $k=3, \dots, 6$ the ground state is the $k$-core since $E_k <0$. At $k=7$ the $k$-core is metastable since $E_7 > 0$.

\begin{table}[!h]
 \caption{$k$-core states in the model Eq. (\ref{eq:4}) at the negative field Eq. (\ref{eq:3}). $M_{k}^{(sim)}$ is the fraction of active agents in the ER network of size $N=10^4$, the mean degree $\langle q \rangle=10$, and the field parameter $\delta =0.001$. $M_{k}^{(theor)}$ is found from the analytical solution \cite{dorogovtsev2006k}. $\langle q \rangle_k$ is the mean degree in the $k$-core, $E_k$ is the energy of the $k$-core from Eq. (\ref{eq:8}).}
 \centering
 \begin{tabular}{c|c|c|c|c|c}
  \hline
  $k$ & 3 & 4 & 5 & 6 & 7 \\
   \hline
   ~$M_{k}^{(sim)}$~ & 0.99722 & 0.98945 & 0.96856 & 0.91756 & 0.74552 \\
   ~$M_{k}^{(theor)}$~ & 0.9971 & 0.98943 & 0.96824 & 0.91781 & 0.74529 \\
   ~$\langle q \rangle_k$~ & 10.018 & 10.05 & 10.094 & 10.01 & 9.7 \\
   ~$E_k$~ & -3,001 & -2,003 & -1,013 & -0,004 & 0.858 \\
   \hline
 \end{tabular}
 \label{table1}
\end{table}

In the case of a scale-free degree distribution $P(q) \propto q^{-\gamma}$ with $2 <\gamma \leq 3$, the mean degree $\langle q \rangle_k$ of nodes in the $k$-core is $k \langle q \rangle /q_0$ where  $q_0$ is the minimal degree \cite{dorogovtsev2006k}. Substituting this result into Eq. (\ref{eq:8}) we find that the $k$-core is the ground states at any $|U|$ and the core index $k$ is given by Eq. (\ref{eq:2}). The energy $E_k$ is negative at $ \langle q \rangle > 2q_0$.

The model Eq. (\ref{eq:4}) is equivalent to the Ising model in a heterogeneous external field. In order to show this, we replace the variable $x_i$ by a spin variable $\sigma_i$,
\begin{equation}
\label{eq:1s}
  x_i=\frac{1}{2}(1+\sigma_i).
\end{equation}
where $\sigma_i=\pm 1$ corresponds to  $x_i=1$ and  $x_i=0$, respectively.
We obtain the Hamiltonian
\begin{equation}
\label{eq:2s}
  E  = - \frac{1}{8} w \sum_{ij}  A_{ij} \sigma_{i} \sigma_{j} - \sum_{i=1}^{N} H_{i} \sigma_{i} +E_0,
\end{equation}
where $E_0$  is a constant and
\begin{equation}
\label{eq:1s}
  H_i=\frac{w}{4}\Bigl(q_i - \frac{2|U|}{w}\Bigr)=\frac{w}{4}[q_i - 2(k-1+\delta)].
\end{equation}
The local field $H_i$ can be either positive or negative depending on degree $q_i$ and  $|U|$.

\section{Structural changes signalling the avalanche collapse}
\label{warnings}

Critical slowing down (decrease of the relaxation rate) is a well-known critical phenomenon, which appears when a system approaches a critical point of both a continuous and discontinuous phase transitions observed in various physical, biological, technological, and social systems. This phenomenon is warning sign of the phase transitions  \cite{stanley1971introduction,scheffer2009early,scheffer2012anticipating,dai2012generic,dai2013slower,dakos2014critical,chialvo2010emergent,lee2014critical,castellano2009statistical,torok2017cascading}.
In this section we show that apart the critical slowing down there are also critical changes in the structure of interaction networks. These changes precede the network collapse. The structural changes create grounds for long-lasting avalanches and  critical slowing down \cite{baxter2015critical}.
They can serve as early warnings of the collapse.

According to \cite{dorogovtsev2006k,dorogovtsev2006kpercolation,goltsev2006k}, nodes of degree $q$ equals to the $k$-core index (i.e., $q=k$) at $k \geq 3$ play a special role in structural stability of the $k$-core. These nodes, which are called `corona' nodes,  form  `corona' clusters inside the $k$-core. If a `corona' node belonging to a `corona' cluster is removed then all other nodes belonging to the same `corona' clusters are also removed  one by one (the domino effect) because their degrees become less than $k$. It is the mechanism of avalanches that destroys the $k$-core at the tipping point \cite{goltsev2006k,baxter2015critical}. These results are valid for $k \geq 3$. The case $k = 2$ corresponds to the ordinary percolation problem.

We introduce a parameter,
\begin{equation}
\label{eq:11}
  \chi_{cr}(k) = \frac{\sum_\alpha s_{\alpha}^2(k)}{\sum_\alpha s_{\alpha}(k)}=\sum_\alpha \pi_\alpha s_{\alpha}(k),
\end{equation}
where $s_{\alpha}(k)$ is the size of a `corona' cluster with index $\alpha$ in the  $k$-core. $\pi_\alpha \equiv s_{\alpha}(k)/\sum_\alpha s_{\alpha}(k)$ is the probability that a randomly chosen corona node in the $k$-core belongs to a corona cluster $\alpha$.
The parameter $\chi_{cr}(k)$ has a meaning of the mean size  of corona clusters to which a randomly chosen corona nodes belongs.
The use of $\chi_{cr}(k)$ can be shown in the case of a randomly damaged network. Random removal of nodes decreases the $k$-core size.
Simultaneously, the number of corona clusters and their sizes increase.
At the critical point of $k$-core collapse the parameter $\chi_{cr}(k)$ diverges in the limit $N \rightarrow \infty$.
Thus, the tipping point of the $k$-core collapse is the percolation point of the `corona' clusters \cite{goltsev2006k}. It is important to note that the growth of corona clusters is the structural mechanism of critical slowing down
when approaching the $k$-core collapse \cite{baxter2015critical}. The parameter $\chi_{cr}(k) $ is similar to the susceptibility, which was introduced in the case of ordinary percolation. Recall that the susceptibility is the mean size  of disjoint clusters to which a randomly chosen node belongs \cite{stauffer1994introduction}.

Based on these results we propose the following method, which allows to reveal structural changes of the interaction network that occur when approaching the tipping point.
For each value of a control parameter, which can be either the field strength, the fraction of removed agents, time,  or temperature, we find $k$-cores by use of the pruning algorithm  and  statistics of corresponding corona clusters by use of the depth-first search algorithm \cite{even2011graph}. Then we calculate the parameter $\chi_{cr}(k) $ from Eq. (\ref{eq:11}). If  $\chi_{cr}(k)$ increases when increasing (or decreasing) the control parameter then it means that the system approaches a point at which the $k$-core disappears. We will apply this method to unweighted and weighted  networks in the next sections.

\section{Structural stability of randomly damaged unweighted networks}
\label{damaged nets}

Let us analyze the structural stability of the model Eq. (\ref{eq:4}) against random damages of the interaction network.
We consider the system of interacting agents  in a heterogeneous negative field $U_i$, which equals to $-(k-1+\delta)$, as well as in Eq. (\ref{eq:3}), with the probability $p$ and  $U_{0} = - q_{cut} -1 \ll - 1$ with the probability $1-p$,  where $q_{cut}$ is the degree cutoff. Thus the probability distribution of $U_i$ is
\begin{equation}
\label{eq:10}
  g(U_{i}) = p\delta(U_{i}+k-1+\delta) + (1-p)\delta(U_{i}-U_{0}).
\end{equation}
If the local field $U_{i}=U_0$ acts on agent $i$ then it deactivates the agent since the strength of interactions is not enough to withhold the agent in active state. Note that agents subjected to this damaging field  are chosen at random with probability $1-p$. The usage of the field Eq. (\ref{eq:10}) with the strong negative component $U_0$ is equivalent to random damage of the interaction network when the fraction $1-p$ of nodes is removed at random.

\begin{figure}[!ht]
 \centering
\includegraphics[width=80mm,angle=0.]{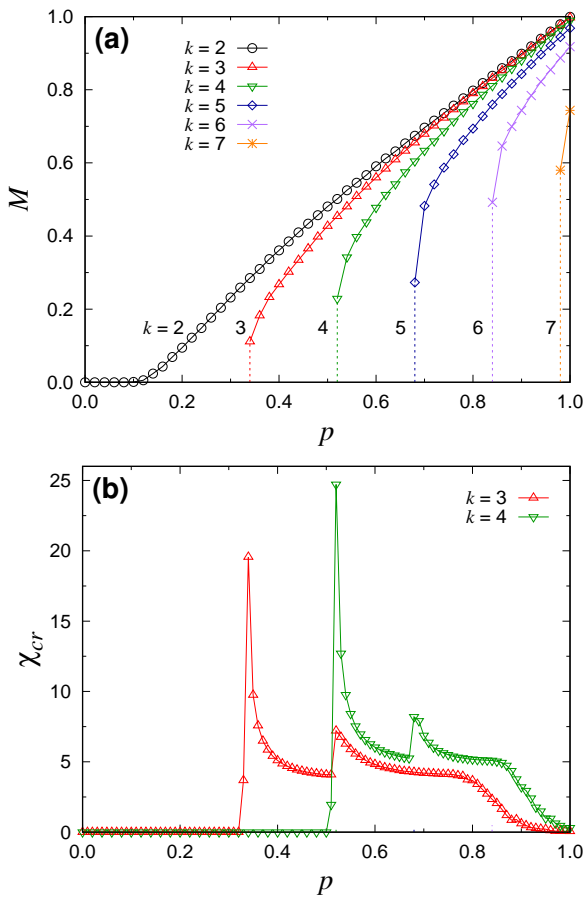}
 \caption{
 %
 (a) Fraction $M$ of active agents versus the occupation probability $p$ in the ground state of a randomly damaged ER network
in the negative field  $U_{i}=-(k-1+\delta)$ at $k=3$ (red triangles) and $k=4$ (green triangles).
(b) The parameter $ \chi_{cr}(k)$ versus $p$ at $k=3$ (red triangles) and $k=4$ (green triangles).
  In simulation we studied ER networks of size $N=10^5$  and the mean degree $\langle q \rangle =10$. Results are averaged over 500 realizations. }
 \label{figMP}
\end{figure}

At first we consider the network of interacting agents when the field is fixed. In general case, applying the pruning algorithm as above, we find that fraction $p$ of remaining active agents forms $k$-core, if it exists. According to \cite{dorogovtsev2006k,dorogovtsev2006kpercolation}, with decreasing $p$,  random damage first destroys the highest core ($k_h $-core). Then it destroys the smaller $(k_h -1)$-core, and so on.  The collapse of the $k$-core with $k \geq 3$ is a hybrid phase transition with a jump of the order parameter as at a first order phase transition, but also with critical fluctuations as at a continuous phase transition. Finally, $2$-core is destroyed and a giant connected component  disappears at the critical point of a continuous phase transition.
This behavior is represented in Figure \ref{figMP}(a), which displays the dependence of the fraction of active agents $M$ in the ground state of the ER network versus the fraction $p$ of randomly chosen agents subjected to the negative field  $U_{i}=-(k-1+\delta)$ at $k=2,3, \dots, 7$ and $\delta=0.001$.
Recall that in this kind of random complex network an edge between each pair of $N$ agents is present with the probability $\langle q \rangle /N$.

Figure \ref{figMP}(b) displays dependence of the parameter $ \chi_{cr}(k)$ on $p$ at the negative fields $U=-(k-1+\delta)$ at $k=3$ and 4, $\delta=0.001$.
$\chi_{cr}(k=3)$ demonstrates a sharp peak at the critical point $p=p_c$ of the $3$-core collapse. Below the critical point, there is no active agent.
It is interesting that $\chi_{cr}(k=3)$ demonstrates one more, but smaller, peak at larger $p$. The second peak corresponds to the collapse of the $4$-core. The origin of the second peak is explained by the fact that corona nodes of $3$-core can be linked with corona nodes of $4$-core. Collapse of corona clusters in $4$-core results in collapse of some corona clusters in $3$-core. Absence of peaks corresponding to collapse of higher $k$-core ($k \geq 5$) can be explained by a small number of corona clusters of $3$-core at large $p$. The parameter $\chi_{cr}(k=4)$ demonstrate a similar behavior in Fig. \ref{figMP}(b). With decreasing $p$ at first a peak of $\chi_{cr}(k=4)$ signals the collapse of $5$-core, then the next sharp peak at smaller $p$ signals the collapse of $4$-core and the whole system.

\section{Structural stability of weighted networks}
\label{weighted nets}

In this section, we consider structural stability of weighted networks of interacting agents  against a uniform negative external field $U$. In the model Eq. (\ref{eq:1}) the weight $w_{ji}$ characterizes the strength of the action of agent $j$ to agent $i$. We introduce the strength $S(i)$ of node $i$   \cite{Barrat04},
\begin{equation}
\label{eq:w1}
S(i)  = \sum_{j} x_{j} w_{ji} A_{ji}.
\end{equation}
It characterizes the force produced by active nearest neighbors of agent $i$ to maintain the agent in the active state. In the case of an unweighted undirected network with $w_{ij}=w_{ji}=1$, the node strength $S(i)$ equals to the number of active nearest neighbors of  agent $i$.

The energy of activation ($x_i =1$) or deactivation  ($x_i =0$) of agent $i$ is
\begin{equation}
\label{eq:w5}
e(i)  = - \Bigl(\sum_{j} x_{j} w_{ji} A_{ji} + U \Bigr)x_i = - [S(i) + U ] x_i.
\end{equation}
If $S(i) > |U|$ then agent $i$ is active in the ground state, otherwise the field deactivates the agent.
We consider the case when weights $w_{ji}$ of edges are uncorrelated random positive numbers with a weight distribution function $G(w)$.
In order to characterize cohesion in weighted networks
we consider a so-called `$S$-weighted' subnetwork as the largest subnetwork  whose nodes  have the node strength at least $S$ \cite{eidsaa2013s}.  The $S$-weighted subnetwork can be found by use of the pruning process, removing one by one all nodes $i$ with the node strength $S(i)$ smaller than $S$.
At a given negative field $U$, the $S$-weighted subnetwork with  $S=|U|$ is the ground state of the model Eq. (\ref{eq:1}).

\begin{figure}[!ht]
 \centering
\includegraphics[width=8cm,angle=0.]{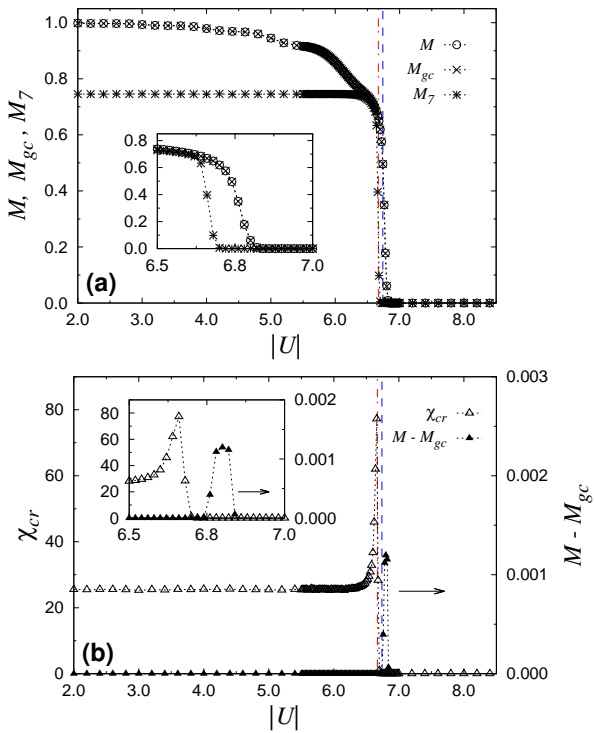}
  \caption{
  (a) ER network of interacting agents with Gaussian weights in a negative field $U$.  $M$ is a fraction of active agents (open circles), $M_7$ is the size of the highest $7$-core (stars), $M_{gc}$ is the size of giant connected component (crosses). Inset represents zoom of the region near the critical point. (b) The $\chi_{cr}(k)$ for $k= 7$ (open triangles) and $M-M_{gc}$ (filled triangles) versus $|U|$.
  The vertical dot-dashed line shows the field at which the highest $k$-core ($k_h = 7$) collapses and $\chi_{cr}(7)$ has a peak. The vertical dashed line is the point $M_{gc}= 0$. Inset represent zoom of the region near the critical point.
    Parameters in simulations: the network size $N=10^4$, the mean degree $\langle q \rangle =10$, the mean weight $\langle w \rangle=1$, the variance $\sigma^2 =0.1$. The number of realizations is 500.}
 \label{figW_gauss}
\end{figure}

In order to understand the interplay of the network topology and the weight distribution function $G(w)$ in the structural stability of the model Eq. (\ref{eq:1}) we use the following numerical methods of network analysis. First, at every $|U|$ we find the $S$-weighted subnetwork.
We use the pruning algorithm for the threshold $S =|U|$. Then, we analyze the topological structure of the $S$-weighted subnetwork  and find the giant connected component of the $S$-weighted subnetwork by use of the depth-first search algorithm \cite{even2011graph}. Furthermore, using the pruning algorithm, we find $k$-cores in the subnetwork and the corona clusters in the $k$-cores by use of the depth-first search algorithm. Then we calculate the parameter $\chi_{cr}(k)$ from Eq. (\ref{eq:11}).

Figure \ref{figW_gauss}(a) displays the dependence of the fraction $M$ of active nodes, the size $M_7$ of the highest core ($k_h=7$), and the parameter $\chi_{cr}(k)$ at $k=7$ versus $|U|$ in the case of the ER network ($\langle q \rangle =10$) with the Gaussian distribution $G(w)$ of weights and the variance $\sigma^2= 0.1$. Random weights smooth the stepped behavior of $M$ in Fig. \ref{MvU}, though
this behavior  is still seen well.  The parameter $\chi_{cr}(k=7)$ in Fig. \ref{figW_gauss}(b) has a sharp peak at a field strength
a little bit smaller than the critical field of the collapse of the whole system.
To understand this result we find the giant connected component $M_{gc}$ of the interaction network of active agents.
Figure \ref{figW_gauss}(a) represents the fraction $M_{gc}$ of active agents in the giant connected component versus $|U|$.
In Fig. \ref{figW_gauss}(b)  we plot the difference $M-M_{gc}$ versus $|U|$. One can see that before the collapse the difference is zero, i.e., $M=M_{gc}$. It means that the active agents form a giant connected component and there is no disjoint cluster of active agents. However, above the point at which $M_{gc}$ disappears ($M_{gc}=0$) there is a narrow region of $|U|$ in which $M-M_{gc}= M \neq 0$. In this region there are only disjoint finite clusters of active agents.
These clusters are formed by strongly interacting agents.
The width of the region between the critical point of the highest core collapse and the critical point of the disappearance of $M_{gc}$ increases when the variance of the weight distribution function $G(w)$ increases.

\begin{figure}[!ht]
 \centering
\includegraphics[width=8cm,angle=0.]{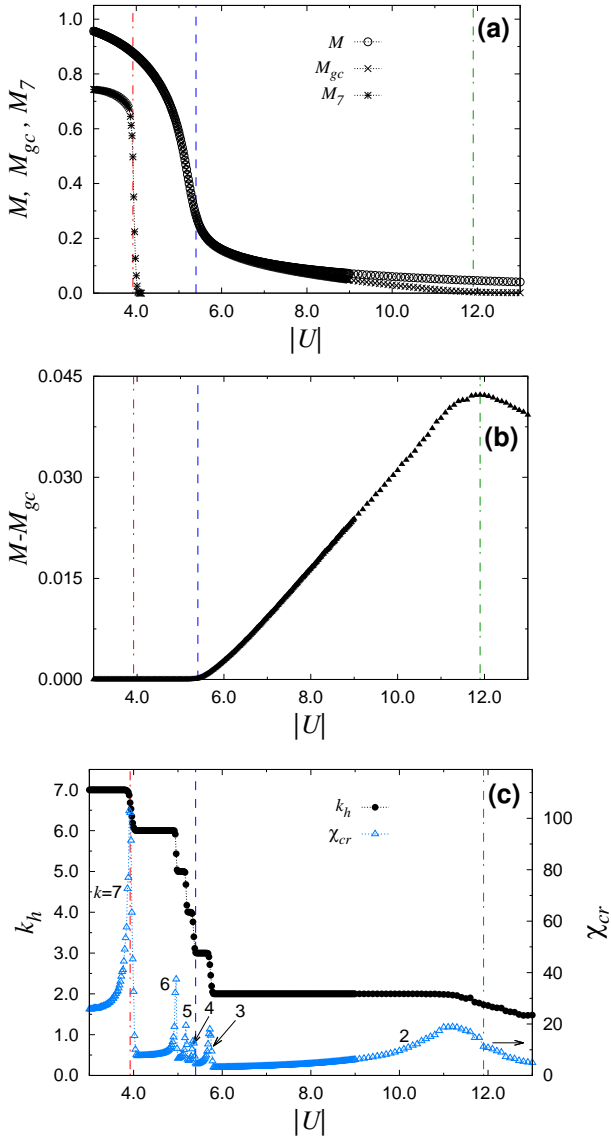}
 \caption{
 (a) The fraction $M$ of active agents, the fraction $M_7$ of the highest $7$-core, and  the fraction $M_{gc}$ of the giant connected component versus $|U|$ in the ER networks with a fat-tailed distribution of weights.
 (b) Field dependence of the fraction $M-M_{gc}$ of finite clusters of active gents.
  (c) The index $k_h$ of the highest $k$-core and $\chi_{cr}(k_h)$ versus the field strength $|U|$. The vertical dash-dotted line on the left corresponds to the critical field of the collapse of the highest $k$-core. The vertical dashed line shows the field above which finite clusters of active agents appear.  The vertical dash-dotted line on the right corresponds to the critical field of disappearance of $M_{gc}$.
  Other parameters: the network size $N=10^5$, the mean degree $\langle q \rangle =10$, the exponent $\alpha =2.5$ of the distribution function of weights. The number of realizations is 500. }
 \label{fig: ER-power}
\end{figure}

\begin{figure}[!ht]
 \centering
 \includegraphics[width=8cm,angle=0.]{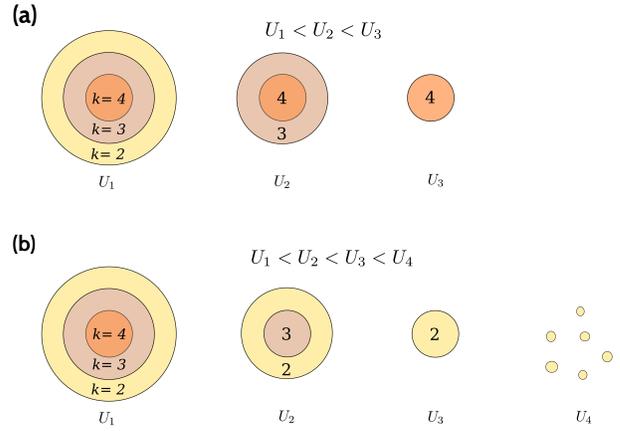}
 \caption{
  Schematic representation of the evolution of the $k$-core organization when increasing the field strength $|U|$: (a) unweighted networks,  $(b)$) weighted  networks. The $k$-cores are represented as nested circles. The largest circle is the 2-core, which includes the higher cores with $k=3,4$. In unweighted networks, when increasing $|U|$ the highest $k$-core disappears last. In the weighted networks  with a fat-tailed distribution of weights, the 2-core disappears last. At sufficiently large fields ($|U| >|U_4|$), only finite disjoint clusters of strongly interacting agents may exist.}
 \label{fig: schematic}
\end{figure}

Let us consider interaction networks with a fat-tailed weight distribution, $G(w)=A w^{-\alpha}$ where $w > 0$ and $ 2 < \alpha \leq 3$. Figures \ref{fig: ER-power} (a)-(c) display results of simulations of the model Eq. (\ref{eq:1}) on the ER network with the power-law weight distribution with $\alpha =2.5$. As one can see in Fig. \ref{fig: ER-power}(a) the  difference between the point of the highest core collapse and the point of the disappearance of $M_{gc}$ is much larger than in the case of the Gaussian distribution of weights in Fig. \ref{figW_gauss}(a).
As one can see in Fig. \ref{fig: ER-power}(b), in a broad range of the fields we have $M=M_{gc}$. It means that the network of active agents consists of only a giant connected component. The finite clusters of active agents appear above a critical point and their fraction is $M-M_{gc} > 0$. Then $M-M_{gc}$ first increases, reaches a maximum at a point at which the giant connected component disappears, i.e., $M_{gc}=0$, and then it decreases.
In order to obtain a detailed information about structural changes in the ground state,  we analyze the $k$-core organization of the state at every $|U|$. Figure \ref{fig: ER-power}(c) represents the dependence of the highest core index $k_h$ and the corresponding parameter $\chi_{cr}(k_h)$  against $|U|$ in the case of $\alpha=2.5$. With increasing $|U|$, the index $k_h$ decreases in a step-like way, reaching the value $2$. The peaks of $\chi_{cr}(k_h)$ point out the field at which the topology of the interaction network is changed. Note that a giant connected component of a network, which has the 2-core as the highest core, has a peculiar topological properties. Namely, nodes of degree $q \geq 2$ in the 2-core are connected by long chains of nodes of degree $q=2$. There are also numerous long branches attached to the core. The giant connected component  disappears at a critical field above which there are only disjoint clusters of active agents. The maximum of $\chi_{cr}(k=2)$ signals this critical continuous transition as one can see in  Fig. \ref{fig: ER-power}(b).
Schematic representation of the evolution of the $k$-core organization of unweighted and weighted interaction networks is given on Fig. \ref{fig: schematic}.

\section{Network stability against thermodynamic fluctuations}
\label{simulation}

Let us study the structural stability of an unweighted network of interacting agents against fluctuations.
We consider the following stochastic process. Agent $i$ transits from a state $x_{i}^{(a)}$ into state $x_{i}^{(b)}$ with a rate $W_{a\rightarrow b}$  determined by the Metropolis algorithm \cite{metropolis1953equation},
\begin{equation}
\label{eq:12}
W_{a\rightarrow b} =
  \begin{cases}
    \tau^{-1} \exp(-\Delta E_{ab} / k_{B}T)~,~~&\Delta E_{ab} > 0~, \\
    \tau^{-1}~,~&\Delta E_{ab} \leq 0 ~,
  \end{cases}
\end{equation}
where
\begin{equation}
\label{eq:13}
\Delta E_{ab} {=}e_{b}(i) {-}e_{a}(i){=}{-} \Bigl(\sum_{j} A_{ij}x_{j} {-} k {+}1 {-}\delta\Bigr)[x_{i}^{(b)} {-}x_{i}^{(a)}],
\end{equation}
$e_{a}(i)$ and $e_{b}(i)$ are the energies Eq. (\ref{eq:5}) of agent $i$ in the states $a$ and $b$. $\tau$ is the time unit for the  update of agent states. If $\Delta E_{ab} \leq 0$, then the update from the state $a$ to the state $b$ is
accepted. If $\Delta E_{ab} > 0$, then a random number $r=[0,1]$ is generated and the agent is updated when $r < \exp(-\Delta E_{ab} /k_{B}T)$. $T$ is the `temperature' of fluctuations. In simulations, the update of agent states is done in parallel, starting from an initial state.

\begin{figure}[!ht]
 \centering
 \includegraphics[width=8cm,angle=0.]{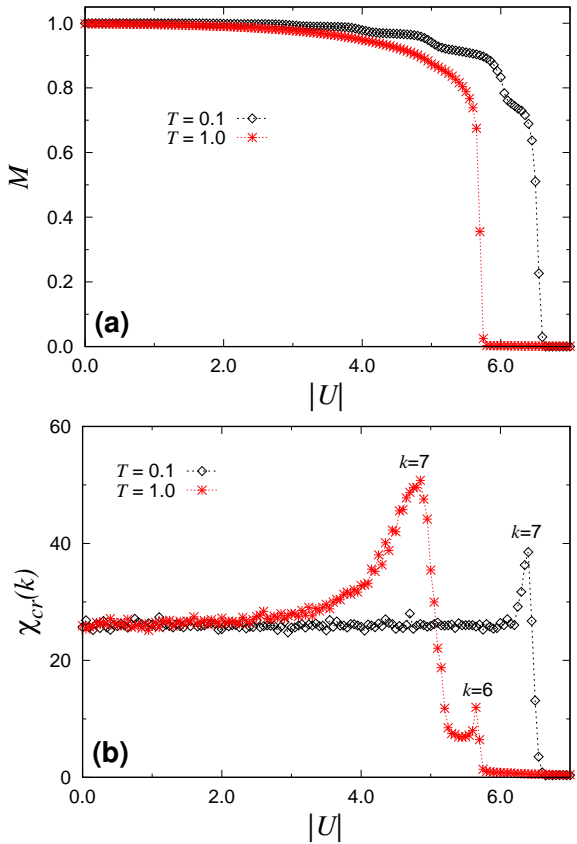}
 \caption{
(a) Fraction of active agents $M$ versus the field strength $|U|$ at the temperatures $T=0.1$ (diamonds) and 1.0 (stars). (b) The parameter  $\chi_{cr}(k)$  versus $|U|$  at $T=0.1$ (diamonds) and  $T=1$  (stars).
The unweighted ER network in Fig. \ref{MvU} is used. The number of realizations is 500. }
 \label{fig MXT}
\end{figure}

Figures \ref{fig MXT}(a) and (b) represent the field dependence of the fraction $M$ of active agents and the parameter $\chi_{cr}(k)$, Eq. (\ref{eq:11}), in the equilibrium state at temperatures $T=0.1$ and 1 in the unweighted ER network of interacting agents as in Sec. \ref{unweighted nets} with the highest $7$-core. One can see that the thermodynamic fluctuations smooth out the stepped behavior of $M$ in Fig. \ref{MvU}. The behavior is still seen well at small temperature $T=0.1$, but almost disappears at $T=1$. This effect is similar to the effect produced by random weights in Fig. \ref{figW_gauss}(a). The critical field $|U_c|$ of the network collapse depends on $T$ and decreases with increasing $T$.
At low $T$, the peak of $\chi_{cr}(k=7)$ as a function of $|U|$ manifests the collapse at $|U_c|$ [see Fig. \ref{fig MXT}(b)]. At higher temperatures, $T=1$, the  negative field first destroys the highest 7-core and then the peak of $\chi_{cr}(k=6)$ signals the collapse of the 6-core and the whole network  of active agents.

\begin{figure}[!ht]
  \centering
\includegraphics[width=8cm,angle=0.]{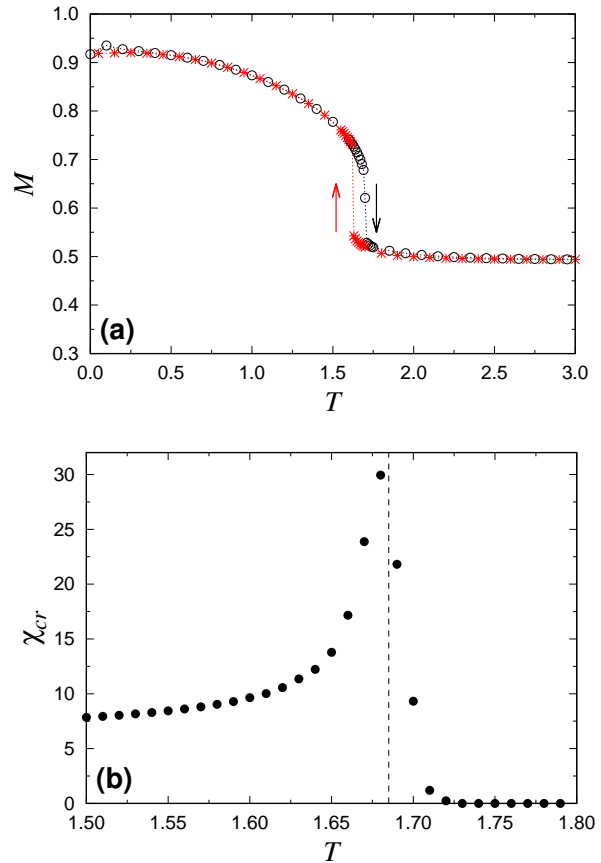}
  \caption{(a) Fraction $M$ of active agents versus temperature $T$ in the unweighted ER network of interacting agents Eq. (\ref{eq:4}) in the uniform negative field $U=-5.001$. Black and red arrows show the directions of the temperature increase and decrease, respectively. (b) The parameter $\chi_{cr}(k=6)$ versus $T$.
  The unweighted ER network
  with the mean degree $\langle q \rangle =10$ and the network size of $N = 10^6$ and $N = 10^5$ were used for (a) and (b), respectively. The number of realizations is 500.
  }
  \label{fig_trans}
\end{figure}

An example of temperature behavior of the ER network of interacting agents at a given field strength $|U|=5.001$ is presented in Fig. \ref{fig_trans}(a). In this field  the 6-core is the ground state at $T=0$ (see Table \ref{table1}). With increasing temperature the network undergoes a first-order phase transition with hysteresis. In order to understand structural changes, which precede the phase transition, we carried out the structural analysis and found corona clusters in the 6-core as described in Sec. \ref{weighted nets}. Figure \ref{fig_trans}(b) shows that the parameter $\chi_{cr}(k=6)$ as  a function of temperature has a sharp peak at the critical temperature $T_c \approx 1.68$. It evidences that the first-order phase transition is driven by the collapse of the 6-core. Above the critical temperature, agents can be active or inactive with almost equal probabilities.

\section{Structural stability of real networks}
\label{real nets}

In this section we apply our model to real networks in ecosystems to analyze the structural stability of this kind of network against external negative factors. 
As an example,
network topology of plants and pollinators in ecosystems is quite well conserved, even though annual variations
of interactions among species are high \cite{BA2011,Alarcon08}. The pollination networks show very specific
 structural property, so-called \textit{nestedness} \cite{Bascompte03}. In the nested networks, \textit{generalists}, which interact
with many other species, play a crucial role for conserving the network stability in contrast to \textit{specialists}, which
prefer to interact with only specific partner \cite{Memmott04}. The \textit{generalists} construct highly connected
subgroups and the whole network can be sustained in a stable state against a decrease of the number of \textit{specialists} unless the
\textit{generalists} are completely removed.

\begin{figure}[!ht]
 \centering
 \includegraphics[width=8.5cm,angle=0.]{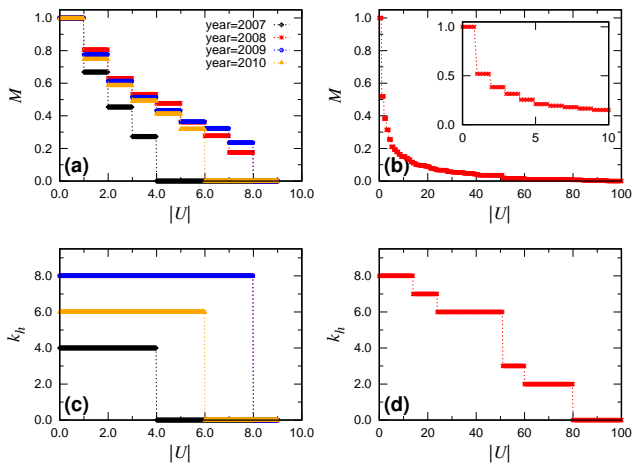}
  \caption{  (a) The fraction $M$ of active agents versus the negative field strength $|U|$ in the unweighted plant-pollinator networks for 4 years \cite{Fang2012}. The numbers of plants-pollinators are (79,126)  at 2007, (88,128) at 2008, (100,165) at 2009, (108,165) at 2010. (b) The fraction $M$ of active agents versus $|U|$ in the plant-fungus network (weighted, 33 plants and 387 fungus) with cutoff level 0.95 of DNA sequence similarity of fungus \cite{toju2014assembly}. The inset represents the zoom of the low field region.
  (c) and (d) represent the index $k_h$ of the highest core versus $|U|$ in the plant-pollinator and plant-fungus networks, respectively.
 }
 \label{figW1}
\end{figure}

As an example of unweighted networks, we use annual observation data for plant-pollinator network in a biodiversity hotspot (Henduan Mountains, Chaina) \cite{Fang2012}. This plant-pollinator network takes only into account  the interactions (visit or not) between species regardless how frequently pollinators visit plants. Our results are displayed in Figs. \ref{figW1} (a) and (c). The number of active agents $M$ decreases with increasing the negative field strength $|U|$ in a step-like way similar to behavior of $M$ in the unweighted ER random network in Fig. \ref{MvU}, but the decrease is faster than in the ER network. We suggest that the nested topology of the pollination network might be a reason of this behavior.
As in the ER network, when the highest $k$-core disappears, all active agents become  inactive [see Fig.\ref{figW1} (a) and (c)].

As an example of weighted networks, we use the below-ground plants-fungus symbioses network (Mt. Yoshida, Kyoto, Japan) \cite{Toju2013,toju2014assembly}.  In the plant-fungus network,
the interactions are assigned with association levels between species. The association level is the number of
root samples in which the focal plant-fungal association was observed \cite{Toju2013}. The plant-fungus networks are slightly different by the
DNA sequence similarity cutoff for fungal taxa, however, the overall network topology is not qualitatively changed.
We used the association level as weights of the plant-fungus interactions.
Unlike other ecological networks, the plant-fungus network has less nestedness and intermediate modularity in comparison with other ecological networks
\cite{toju2014assembly}. The fraction $M$ of active agents density shows still a step-wise decrease [see Fig.\ref{figW1} (b)]. In contrast to the ER random network on Fig. \ref{fig: ER-power},
the parameter $M$ decreases very rapidly, even though the index $k_h$ of the highest $k$-core decreases slowly [see Fig. \ref{figW1} (d)].
We suggest that the fast decrease of $M$  is caused by correlations between the topology of the network and the weights of edges. Assuming that all agents are active in Eq. (\ref{eq:w1}), we find that in the plant-fungus network the averaged strength $s(q)$ of nodes  with degree $q$ is a power law, $s(q)\sim q^{\beta}$, with the exponent $\beta \simeq 1.5$. This kind of the node strength distribution was also found in the world-wide airport network \cite{Barrat04} as a result of correlations between  degree $q_i$ of node $i$ and weights $w_{ji}$ of incoming edges. Note that in uncorrelated networks  $\beta=1$, i.e., $s(q) \propto q$.
%
Thus, small degree agents, which are dominant in the network, have on average a small node strength Eq. (\ref{eq:w1}) and are removed at small field strengths. The fraction of highly connected nodes is very small, only 7$ \%$.
But it is these nodes that form $k$-cores with $k \geq2$ and remain being active at strong negative fields.


\section{Conclusions}

In this work, we studied structural stability of weighted and unweighted networks of positively interacting agents against a negative external field. We showed that positively interacting agents support the activity of each other and confront the negative field, which aims to suppress the activity of the agents. In our approach we understand structural stability as the existence of a giant connected component of the network of active agents stable against perturbations. The competition between positive interaction and the negative field shapes the structure of stable states of the networks.
In a network with a uniform interaction between agents (unweighted networks), the ground state of active agents has the structure of $k$-core with  the core index $k$ determined by the field strength. With increasing the field strength the network of active agents undergoes a cascade of transitions from $k$-core to $(k+1)$-core ground state.
The field  destroys first $2$-core, then $3$-core, and so on. The highest $k_h$-core is destroyed at last. There is a critical field strength (tipping point) above which the system collapses into a state with inactive agents. The critical  point is determined by the highest $k$-core.
In contrast, increasing random damage (removal of agents at random) destroys at first the highest $k_h$-core, then $(k_h -1)$-core, and so on. $2$-core is destroyed at last.

Weighted networks of interacting agents (networks with heterogeneous interactions) in a negative external field demonstrate a behavior opposite to unweighted networks but similar to networks under random damage.
In the case of a sufficiently narrow weight distribution, increasing the negative field strength destroys at first the highest $k_h$-core, then $(k_h -1)$-core, up to a certain critical $k_c$-core determined by the network  structure and weight distribution. In a weighted network  with a fat-tailed distribution of weights, increasing  field strength destroys first the highest $k$-core and 2-core disappears last. At sufficiently large fields, only finite disjoint clusters of active strongly interacting agents may exist in this case. Thus, networks of interacting agents with fat-tailed distributions of interaction strengths and degrees are robust against both a negative external field and random damage. Namely, there is a finite fraction of active agents forming a giant connected cluster which confronts even very strong negative external factors.


In this paper, we also demonstrated that a critical change in the structure of the system precedes the $k$-core collapse. It is signalling the approach to the tipping point. These structural changes create grounds for long-lasting avalanches and  critical slowing down. They can serve as early warnings of the collapse. We proposed a new method of structural analysis that allows to reveal structural changes caused by external forces in the unweighted and weighted interaction networks. We showed that an analysis of $k$-core organization and  statistics of so-called `corona' clusters give a powerful tool to  investigate evolution of structure in an external negative field or under damage. This method allows to predict collapse of $k$-cores. The structural changes are caused by the growth of clusters of corona nodes in $k$-core.  At the critical point, the clusters percolate. As a result, the second moment of the size distribution of the corona clusters in the core diverges in the thermodynamic limit.
We apply this method to unweighted and weighted networks of interacting agents. For every value of a control parameter, which can be either the field strength, the fraction of removed agents, time,  or temperature, we found $k$-cores of the network of active agents,  statistics of corona clusters in the $k$-cores,  and the parameter $\chi_{cr}(k) $ from Eq. (\ref{eq:11}). If  $\chi_{cr}(k)$ increases when increasing (or decreasing) the control parameter then it means that the system approaches a point at which the $k$-core collapses.

This work was focused on the structural stability of networks of positively interacting agents.
One can consider the case when there are both positive (mutualistic) and negative (antagonistic) interactions. This model can be applied to study the structural stability of ecosystems with antagonistic interactions \cite{toju2014assembly,mougi2012diversity}.

Another interesting case is to study interacting agents in a non-uniform negative external field. This case needs a consideration of heterogeneous $k$-cores \cite{baxter2011heterogeneous,cellai2011tricritical}. One can show that the case, when  the external field can be both positive and negative, corresponds to bootstrap percolation problem \cite{baxter2011heterogeneous,baxter2010bootstrap}. In this case, agents in positive local field are always active, in other words, they are seeds of activation.

In Sec. \ref{unweighted nets}  we demonstrated the equivalence of the network of interacting agents to a network of interacting Ising spins in a non-uniform magnetic field. If positive and negative interactions are distributed randomly, then the network of interacting agents is equivalent to the Ising spin glass model on a network in a non-uniform magnetic field.

In this work we also studied structural stability of  interacting agents against thermal fluctuations. At finite temperatures, we analyzed  the equilibrium state by use of the Metropolis algorithm. However, one can also study dynamics based on other rules like ones in the investigations of ecosystems or social networks.

\section{Acknowledgements}
This work was supported by
the grant PEST UID/CTM/50025/2013. S.Y. acknowledges the financial support of Instituto de Nanoestruturas, Nanomodela‹o e Nanofabrica‹o via grant BPD/UI96/5443/2016.

\bibliography{mybib_kcore_state}

\end{document}